\documentclass[
aps,
pra,
reprint,
notitlepage,
superscriptaddress,
longbibliography
]{revtex4-2}

\usepackage[T1]{fontenc}
\usepackage{amsmath,amssymb,amsthm,amsfonts}
\usepackage{bm}
\usepackage{physics}
\usepackage{mathtools}
\usepackage{graphicx}
\usepackage{xcolor}
\usepackage{hyperref}
\usepackage{float}
\usepackage{booktabs}
\usepackage{multirow}
\usepackage{subcaption}
\usepackage{algorithm}
\usepackage{algpseudocode}

\theoremstyle{plain}

\theoremstyle{definition}

\theoremstyle{remark}

\begin{document}

\title{Closed-Form Expectation Values of the Damped Kerr Oscillator}

\author{Jacob Emerson}
\affiliation{Department of Electrical and Computer Engineering, Princeton University}

\date{\today}

\begin{abstract}

We derive a closed-form analytical expression for the expectation values of a damped Kerr nonlinear oscillator initialized in a coherent state. Starting from the exact Liouville-space solution of the Lindblad master equation, we specialize to coherent-state initial conditions, obtaining explicit time-dependent expressions for arbitrary normally ordered expectation values. The resulting expressions depend only on the system parameters and the initial coherent-state amplitude, requiring neither numerical integration nor evolution in a truncated Fock space. We verify the analytical expressions against master-equation simulations over a range of nonlinear interaction strengths. The closed-form solution provides an efficient alternative to numerical simulation while remaining free of Fock-space truncation error. Finally, we note an extension of the underlying Lie-algebraic structure that may provide a useful starting point for obtaining exact solutions to a broader class of open quantum optical systems.

\end{abstract}

\maketitle

% ============================================================
\section{Introduction}
Open quantum systems with Kerr nonlinearities provide a canonical model for nonlinear quantum optics and have become an important platform in superconducting circuit quantum electrodynamics \cite{ScullyZubairy,GerryKnight2005,BlaisReview}. A prototypical example is the Kerr nonlinear oscillator, described by the Hamiltonian
\begin{equation}
H=\omega a^\dagger a+\chi(a^\dagger a)^2,
\end{equation}
which arises naturally in nonlinear optical media and in superconducting microwave resonators through Josephson nonlinearities \cite{ScullyZubairy,GerryKnight2005,BlaisReview,Krantz2019}. Although the Hamiltonian has a simple form, the quartic Kerr interaction generates non-Gaussian dynamics that generally preclude Gaussian solution techniques and make both analytical calculations and numerical simulations substantially more challenging than for harmonic systems.

The nonlinear dynamics generated by the Kerr interaction give rise to phenomena including phase-space shearing, collapse and revival of coherent states, and the formation of Schr\"odinger-cat states \cite{YurkeStoler1986,Milburn1986SchrodingerCats,GerryKnight2005}. These nonclassical effects have made Kerr systems an important platform for continuous-variable quantum information processing and for investigating decoherence in nonlinear quantum systems.

In realistic settings, Kerr oscillators are inevitably coupled to their environment, leading to dissipative dynamics described by a Lindblad master equation. While exact propagators for the damped Kerr oscillator have been derived using Lie-algebraic and thermo-field techniques \cite{Chaturvedi1991}, the resulting solutions are formulated in Liouville space and do not directly yield closed-form expressions for experimentally relevant expectation values. Consequently, numerical integration in a truncated Fock space remains the standard approach despite its rapidly increasing computational cost with Hilbert-space dimension.

In this work, we derive closed-form analytical expressions for arbitrary normally ordered moments
\[
\langle (a^\dagger)^\ell a^k\rangle
\]
of a damped Kerr oscillator initialized in a coherent state. Our derivation exploits the underlying $\mathfrak{su}(1,1)$ structure of the Liouvillian to evaluate the coherent-state matrix elements analytically, yielding explicit expressions for arbitrary normally ordered observables. Quantities such as the field quadrature expectation value $\langle X(t)\rangle$ and higher-order moments such as $\langle P^2(t)\rangle$ follow immediately as special cases.

Unless otherwise stated, we work in natural units with $\hbar=1$.

\section{Theoretical Framework}
\label{sec:theoreticalFramework}

Under the standard Born--Markov and secular approximations, the dynamics of an open quantum system are governed by the Lindblad master equation \cite{BreuerPetruccione,Lindblad1976},
\begin{equation}
\begin{split}
\dot{\rho}
=
\mathcal{L}[\rho]
=
-i[H,\rho]
+\kappa\mathcal{D}[a]\rho,
\\
\mathcal{D}[a]\rho
=
a\rho a^\dagger
-\frac12
\left(
a^\dagger a\rho
+\rho a^\dagger a
\right).
\end{split}
\label{eq:ME}
\end{equation}
Here $a$ and $a^\dagger$ satisfy the canonical commutation relation
$[a,a^\dagger]=1$, and $\hat n=a^\dagger a$ is the number operator. We consider the zero-temperature limit relevant to many superconducting microwave resonator experiments, for which photon loss is the dominant dissipative process.

We consider the Kerr oscillator Hamiltonian,
\[
H=\omega\hat n+\chi\hat n^2.
\]
A common alternative convention is
\[
H=\omega'\hat n+\frac{K}{2}\hat n(\hat n-1),
\]
which is equivalent under the parameter redefinitions
\[
\chi=\frac{K}{2},
\qquad
\omega=\omega'-\frac{K}{2}.
\]
Throughout this work we adopt the former convention for notational simplicity.

Before considering dissipation, it is useful to recall that the undamped Kerr oscillator is exactly solvable for coherent initial states owing to three key features: (i) the Hamiltonian is diagonal in the Fock basis, $H=f(\hat n)$; (ii) the photon-number probabilities of a coherent state follow a Poisson distribution; and (iii) the transition energies,
\[
E_{n+1}-E_n=\omega+\chi(2n+1),
\]
depend linearly on the occupation number $n$. These properties permit many coherent-state expectation values to be evaluated analytically and motivate the approach developed here for the dissipative case.

These properties lead to the well-known coherent-state result
\begin{equation}
\langle a(t)\rangle
=
\alpha e^{-i(\omega+\chi)t}
\exp\!\left[|\alpha|^2\left(e^{-2i\chi t}-1\right)\right].
\end{equation}
This solvability is non-generic: most nonlinear Hamiltonians do not admit such closed forms because the moment hierarchy generated by
\[
\frac{d}{dt}\langle a\rangle
=
-i\omega\langle a\rangle
-i\chi\langle(\hat n-1)a\rangle
\]
does not close, producing an infinite chain of coupled moments
\[
\langle a\rangle
\rightarrow
\langle \hat n a\rangle
\rightarrow
\langle \hat n^2 a\rangle
\rightarrow
\cdots.
\]
The Kerr oscillator is therefore unusual in that this otherwise unclosed hierarchy can be resummed exactly for coherent initial states.

To exploit the Lie-algebraic structure of the Liouvillian, we adopt the vectorized (thermo-field) representation of \cite{Chaturvedi1991}, in which the density matrix is identified with a vector in a doubled Hilbert space. The density matrix
\[
\rho=\sum_{m,n}\rho_{mn}|m\rangle\langle n|
\]
is mapped to
\[
|\rho\rangle
=
\sum_{m,n}\rho_{mn}|m,n\rangle,
\qquad
|m,n\rangle
\equiv
|m\rangle\otimes|n\rangle,
\]
under which the master equation takes the form
\[
\frac{d}{dt}|\rho\rangle=\mathcal{L}|\rho\rangle,
\]
with the Liouvillian acting as an operator on the doubled Hilbert space.
We introduce left and right bosonic operators acting on each tensor factor:
\[
a_L|m,n\rangle=\sqrt{m}\,|m-1,n\rangle,\quad
a_L^\dagger|m,n\rangle=\sqrt{m+1}\,|m+1,n\rangle,
\]
\[
a_R|m,n\rangle=\sqrt{n}\,|m,n-1\rangle,\quad
a_R^\dagger|m,n\rangle=\sqrt{n+1}\,|m,n+1\rangle,
\]
which satisfy
\[
[a_L,a_L^\dagger]=1,\qquad
[a_R,a_R^\dagger]=1,\qquad
[a_L^{(\dagger)},a_R^{(\dagger)}]=0.
\]
Left and right multiplication of operators on $\rho$ therefore correspond to actions on the first and second tensor factors, respectively. This formulation exposes the Lie-algebraic structure of the Liouvillian, enabling the disentangling procedure used to derive the exact propagator in the following section.
%============================================================
\section{Liouvillian Disentangling}
\label{sec:decomposition}
%============================================================
We now derive the exact propagator for the damped Kerr master equation using the vectorized $\mathfrak{su}(1,1)$ formulation of \cite{Chaturvedi1991}, which serves as the foundation for the closed-form evaluation of coherent-state moments.

The Liouvillian can be expressed in terms of generators of the $\mathfrak{su}(1,1)$ algebra,
\[
K_+=a_L^\dagger a_R,\qquad
K_-=a_La_R^\dagger,\qquad
K_3=\tfrac12(a_L^\dagger a_L+a_R^\dagger a_R+1),
\]
together with the central operator
\[
K_0=a_L^\dagger a_L-a_R^\dagger a_R.
\]
These satisfy
\[
\begin{aligned}
[K_-,K_+] &= 2K_3,\quad [K_3,K_\pm] = \pm K_\pm,\\
[K_0,K_\pm] &= [K_0,K_3] = 0.
\end{aligned}
\]
In terms of these operators, the Kerr Liouvillian becomes
\begin{equation}
\mathcal L
=
-i\omega K_0
-i\chi K_0(2K_3-1)
+\kappa K_-
-\kappa K_3
+\frac{\kappa}{2}.
\label{eq:liouvillian}
\end{equation}
Since $K_0$ commutes with the $\mathfrak{su}(1,1)$ generators, the propagator factorizes as
\[
e^{\mathcal Lt}
=
e^{(\frac{\kappa}{2}-i(\omega-\chi)K_0)t}
U(t),
\]
where the remaining evolution operator is disentangled using the Wei--Norman factorization \cite{WeiNorman1963},
\[
U(t)
=
e^{\Gamma_-(t)K_-}
e^{\Gamma_3(t)K_3}.
\]
Substituting this ansatz into the evolution equation and applying the Baker--Campbell--Hausdorff identities (see Appendix~\ref{app:WeiNorman}) yields
\[
\dot{\Gamma}_3=\lambda,\qquad
\dot{\Gamma}_-+\lambda\Gamma_-=\kappa,
\qquad
\lambda=-2i\chi K_0-\kappa,
\]
with initial conditions
\[
\Gamma_3(0)=0,\qquad
\Gamma_-(0)=0,
\]
required by $U(0)=I$. Solving these equations gives
\[
\Gamma_3=\lambda t,\qquad
\Gamma_-=
-\frac{\kappa}{\lambda}
\left(e^{-\lambda t}-1\right),
\]
so that
\begin{equation}
e^{\mathcal Lt}
=
e^{(\frac{\kappa}{2}-i(\omega-\chi)K_0)t}
e^{-\frac{\kappa}{\lambda}(e^{-\lambda t}-1)K_-}
e^{\lambda tK_3}.
\label{eq:propfactorized}
\end{equation}

Applying the disentangled propagator Eq. (\ref{eq:propfactorized}) to the Liouville Fock basis gives
\begin{equation}
\begin{aligned}
e^{\mathcal Lt}|m,n\rangle
&=
e^{(\frac{\kappa}{2}-i(\omega-\chi)(m-n))t}
e^{\frac{\lambda}{2}(m+n+1)t}
\\
&\times \sum_{r=0}^{\min(m,n)}
\frac{
\left[
-\frac{\kappa}{\lambda}
\left(e^{-\lambda t}-1\right)
\right]^r
}{r!}
\\
&\quad \times
\sqrt{
\frac{m!n!}
{(m-r)!(n-r)!}
}
\,|m-r,n-r\rangle,
\end{aligned}
\label{eq:propagator}
\end{equation}
where $\lambda=-2i\chi(m-n)-\kappa$. The propagator in Eq.~(\ref{eq:propagator}) is completely general and therefore applies to arbitrary initial density operators. The coherent-state assumption enters only in the analytical evaluation of the remaining double sums in the following section. Consequently, the present procedure extends directly to other initial states; the existence of a closed-form solution is then determined by the structure of the corresponding Fock-space expansion rather than by the propagator itself.
%===================================================
\section{Closed-form Evaluation of Normally Ordered Moments}
%===================================================
Having obtained the general propagator, we now specialize to coherent initial states, for which the remaining sums can be evaluated in closed form. In this case, the vectorized density operator obeys

\begin{equation}
|\rho(0)\rangle
=
e^{-|\alpha|^2}
\sum_{m,n=0}^{\infty}
\frac{\alpha^m(\alpha^*)^n}
{\sqrt{m!n!}}
|m,n\rangle .
\label{eq:coherentState}
\end{equation}
Applying the propagator of Eq.~(\ref{eq:propagator}) term by term to the coherent-state expansion yields

\begin{equation}
\begin{aligned}
|\rho(t)\rangle
&=
e^{-|\alpha|^2}
\sum_{m,n=0}^{\infty}
\frac{\alpha^m(\alpha^*)^n}{\sqrt{m!n!}}
e^{(\frac{\kappa}{2}-i(\omega-\chi)(m-n))t}
\\
&\times
e^{\frac{\lambda}{2}(m+n+1)t}\sum_{r=0}^{\min(m,n)}
\frac{\Gamma_-^r}{r!}
\\
&\quad \times
\sqrt{
\frac{m!n!}
{(m-r)!(n-r)!}
}
|m-r,n-r\rangle .
\end{aligned}
\label{eq:rhot}
\end{equation}
where $\Gamma_-=
-\frac{\kappa}{\lambda}
\left(e^{-\lambda t}-1\right)$. Taking the expectation value
\[
\langle (a^\dagger)^\ell a^k\rangle
=
\operatorname{Tr}\!\left[(a^\dagger)^\ell a^k\rho(t)\right]
\]
using Eq.~(\ref{eq:rhot}) yields

\begin{equation}
\begin{aligned}
\langle (a^\dagger)^\ell a^k\rangle
&= e^{-|\alpha|^2}
\sum_{m,n}
\sum_r
\frac{\alpha^m(\alpha^*)^n}{\sqrt{m!n!}}
\frac{\Gamma_-^r}{r!}
\\
&\times
e^{(\frac{\kappa}{2}-i(\omega-\chi)(m-n))t}
\\
&\quad \times
e^{\frac{\lambda}{2}(m+n+1)t}
\langle n-r|
(a^\dagger)^\ell a^k
|m-r\rangle .
\end{aligned}
\label{eq:momentExpansion}
\end{equation}
From this point onward, $\lambda$ denotes the scalar eigenvalue obtained by evaluating $K_0$ on $|m,n\rangle$. To avoid separate case distinctions, we temporarily assume $k\ge\ell$. Using the standard Fock-space action of the ladder operators, the matrix element in Eq.~(\ref{eq:momentExpansion}) evaluates to
\begin{equation}
\begin{aligned}
\langle n-r|
(a^\dagger)^\ell a^k
|m-r\rangle
=
\sqrt{\frac{(m-r)!}{(m-r-k)!}}
\\
\times
\sqrt{\frac{(m-r-k+\ell)!}{(m-r-k)!}}
\,
\delta_{n-r,m-r-k+\ell}.
\end{aligned}
\label{eq:matrixelt}
\end{equation}
Using Eq.~(\ref{eq:matrixelt}), the Kronecker delta fixes $n=m-k+\ell$, reducing the expectation value to
\begin{equation}
\begin{aligned}
\langle (a^\dagger)^\ell a^k\rangle
&=
e^{-|\alpha|^2}
e^{(\frac{\kappa}{2}
-i(\omega-\chi)(k-\ell))t}
\\
&
\times
\sum_{m=k}^{\infty}
\frac{\alpha^m(\alpha^*)^{m-k+\ell}}
{\sqrt{m!(m-k+\ell)!}}
e^{\frac{\lambda t}{2}(2m-k+\ell+1)}
\\
&
\times
\sum_{r=0}^{m-k+\ell}
\frac{\Gamma_-^r}{r!}
\sqrt{
\frac{m!(m-k+\ell)!}
{(m-r)!(m-r-k+\ell)!}
}
\\
&
\times
\sqrt{
\frac{(m-r)!}
{(m-r-k)!}
}
\sqrt{
\frac{(m-r-k+\ell)!}
{(m-r-k)!}
}.
\end{aligned}
\label{eq:momentReduced}
\end{equation}
Simplifying the factorial prefactors reduces the $r$-sum in Eq.~(\ref{eq:momentReduced}) to
\[
\sum_{r=0}^{m-k+\ell}
\Gamma_-^r
\binom{m-k+\ell}{r}
\frac{(m-r-k+\ell)!}
{(m-r-k)!}
\sqrt{\frac{m!}{(m-k+\ell)!}}.
\]
The remaining square-root factor cancels the coherent-state normalization, and the finite sum is evaluated using the combinatorial identity
\[
\sum_{r=0}^{m}
a^r
\binom{m}{r}
\frac{(m-r)!}
{(m-r-\ell)!}
=
\frac{m!}{(m-\ell)!}
(1+a)^{m-\ell},
\]
whose proof is given in Appendix~\ref{app:identity}.
Applying this identity and the index shift $m\mapsto m+k$ yields
\begin{equation}
\begin{aligned}
\alpha^\ell(\alpha^*)^k
e^{-|\alpha|^2}
&e^{(\frac{\kappa}{2}
-i(\omega-\chi)(k-\ell))t}
\sum_{m=0}^{\infty}
\frac{|\alpha|^{2m}}{m!}
\\
&\times
e^{\frac{\lambda t}{2}(2m+k+\ell+1)}
(1+\Gamma_-)^m.
\end{aligned}
\label{eq:momentGenerating}
\end{equation}
Recognizing the remaining series as the exponential generating function gives the main result Eq.~(\ref{eq:moment}).

\begin{equation}
\begin{aligned}
\langle (a^\dagger)^\ell a^k\rangle
&=
\alpha^\ell(\alpha^*)^k
e^{-|\alpha|^2}
\\
&\times
e^{(\frac{\kappa}{2}
-i(\omega-\chi)(k-\ell))t}
e^{\frac{\lambda t}{2}(k+\ell+1)}
\\
&\quad\times
\exp\!\left[
|\alpha|^2 e^{\lambda t}
\left(
1-\frac{\kappa}{\lambda}
(e^{-\lambda t}-1)
\right)
\right].
\end{aligned}
\label{eq:moment}
\end{equation}
where $\lambda=-2i\chi(k-\ell)-\kappa$

Although the intermediate derivation assumes $k \ge \ell$ to avoid separate case distinctions, Eq.~(\ref{eq:moment}) is valid for arbitrary non-negative integers $k$ and $\ell$. Interchanging $k$ and $\ell$ yields the complex-conjugate expression, as expected from Hermiticity.
As immediate examples, the field quadrature expectation values
\[
\langle X(t)\rangle
=
\frac{\langle a(t)\rangle+\langle a^\dagger(t)\rangle}{\sqrt2},
\qquad
\langle P(t)\rangle
=
\frac{\langle a(t)\rangle-\langle a^\dagger(t)\rangle}{i\sqrt2},
\]
follow from $(k,\ell)=(1,0)$ and $(0,1)$, while other higher-order moments follow directly from Eq.~(\ref{eq:moment}).
%============================================================
\section{Numerical Validation}
\label{sec:validation}
%============================================================

To validate the analytical results, we compare the closed-form expressions with direct numerical integration of the Lindblad master equation using QuTiP. Specifically, we compare the quadrature dynamics over a range of Kerr nonlinearities, examine the numerical error throughout the evolution, compare the computational cost of both approaches, and verify the analytical expression for higher-order normally ordered moments. All parameters are expressed in angular frequency units.
\begin{figure*}[t]
\centering
\includegraphics[width=\textwidth]{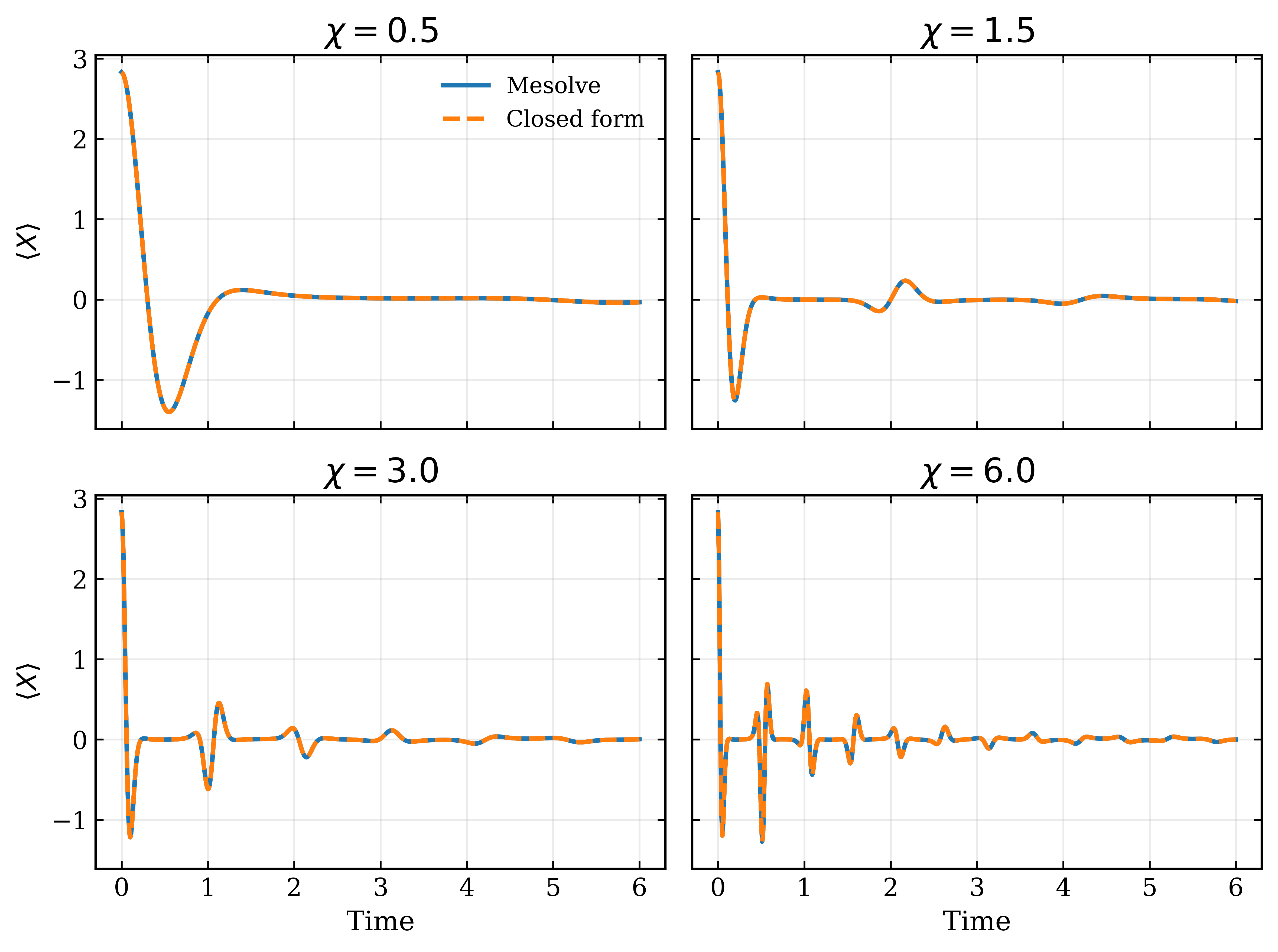}
\caption{
Comparison of the analytical solution and QuTiP master-equation simulation for the quadrature expectation value $\langle X(t)\rangle$ at representative Kerr nonlinearities $\chi$. The analytical and numerical curves are visually indistinguishable across the parameter range considered.
}
\label{fig:kerr}
\end{figure*}
\begin{figure}[t]
\centering
\includegraphics[width=0.76\columnwidth]{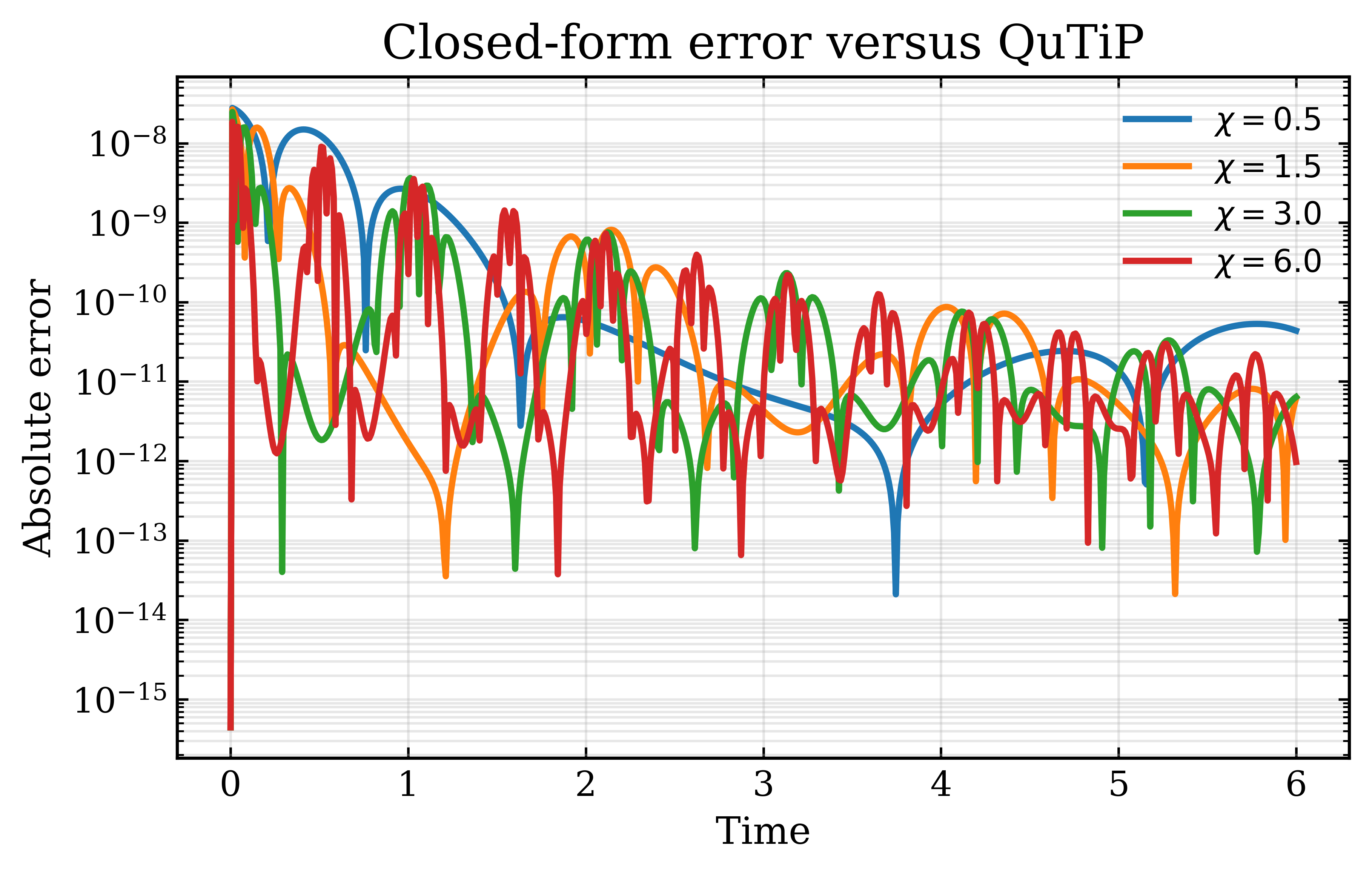}
\caption{
Absolute error between the analytical expression and numerical master-equation simulation for the same parameter values shown in Fig.~\ref{fig:kerr}. The error remains near numerical precision throughout the evolution, confirming the correctness of the closed-form expression (Eq.~(\ref{eq:moment})).
}
\label{fig:error}
\end{figure}
\begin{figure}[t]
\centering
\includegraphics[width=0.85\columnwidth]{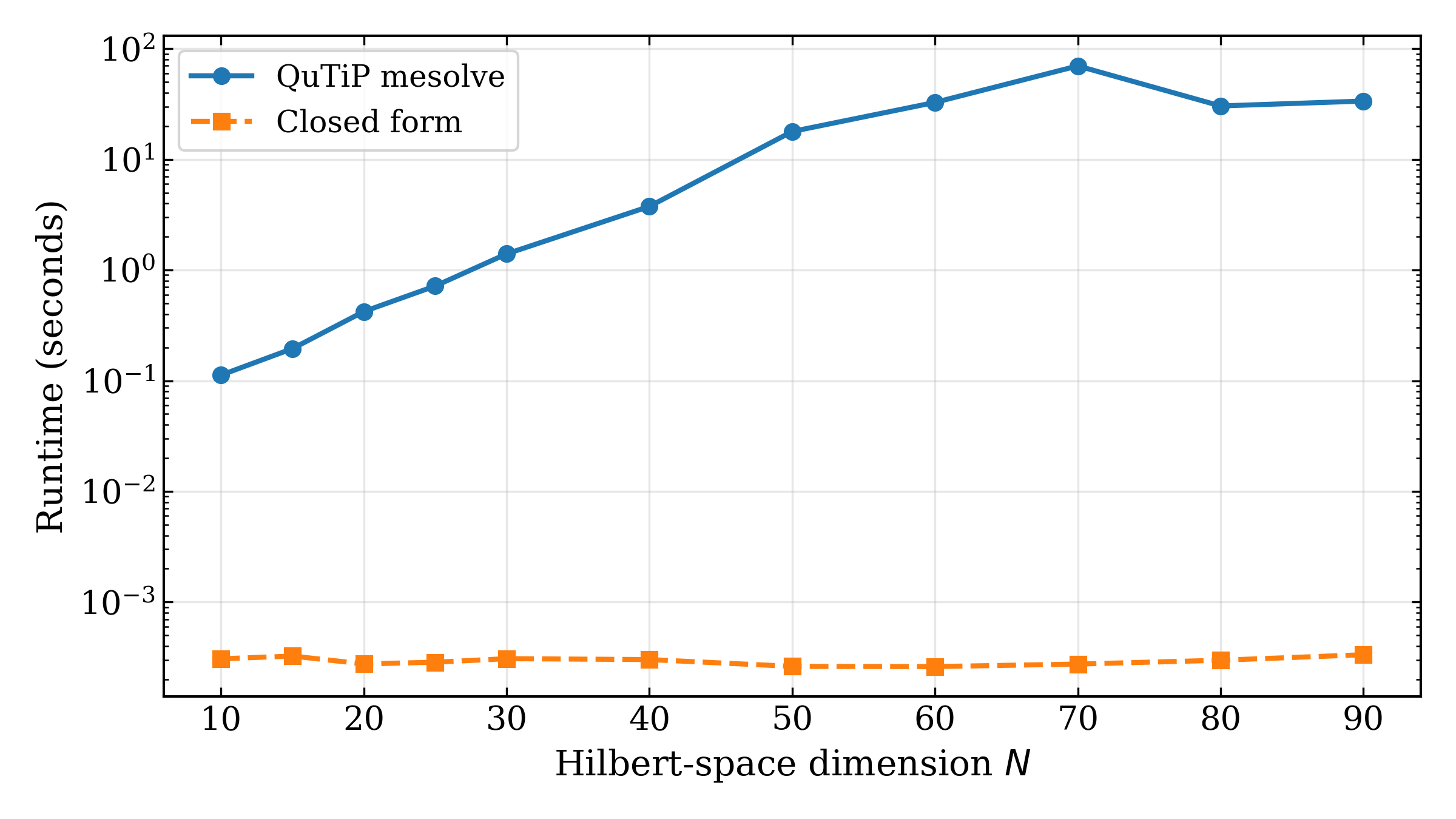}
\caption{
Runtime comparison between evaluation of the analytical expression and numerical integration of the Lindblad master equation as a function of Hilbert-space truncation dimension. While the cost of the numerical solver increases rapidly with the Fock cutoff, the evaluation time of the analytical expression remains essentially constant.
}
\label{fig:runtime}
\end{figure}
\begin{figure*}[t]
\centering
\includegraphics[width=\textwidth]{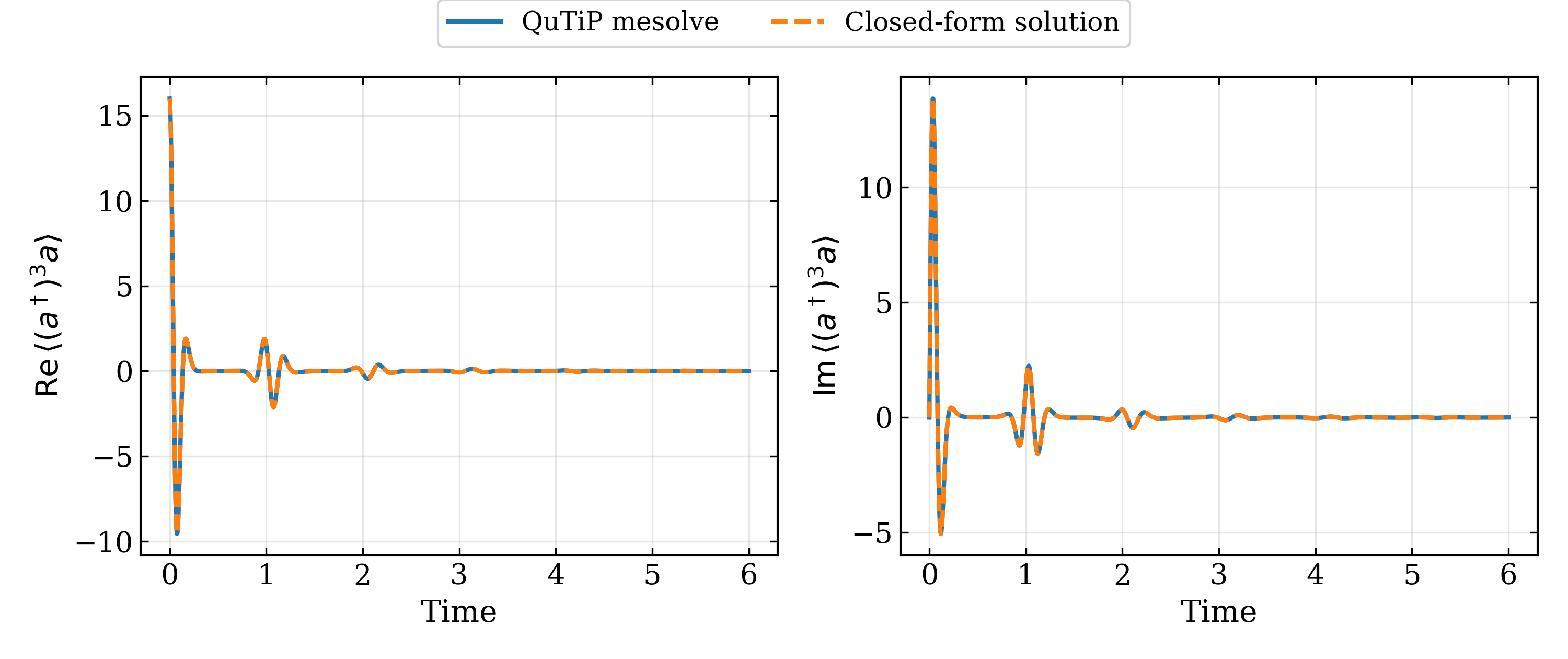}
\caption{
Validation of the analytical solution for the higher-order normally ordered moment $\langle (a^\dagger)^3a\rangle$. The left and right panels show the real and imaginary components, respectively. Excellent agreement with the numerical master-equation solution demonstrates that the closed-form expression (Eq.~(\ref{eq:moment})) applies beyond first-order moments to arbitrary normally ordered operator products.
}
\label{fig:higher}
\end{figure*}

\section{Discussion and Conclusion}
\label{sec:discussion}

The dissipative Kerr oscillator retains a surprising degree of analytical structure when initialized in a coherent state. Although the Kerr interaction generates an infinite hierarchy of coupled moments, the combination of the vectorized $\mathfrak{su}(1,1)$ Liouvillian representation and the Poisson structure of coherent states reorganizes the resulting expressions into sums that admit exact resummation. The resulting solvability therefore arises not from moment closure, but from the compatibility between the Liouvillian algebra and the coherent-state expansion.

From a practical perspective, Eq.~(\ref{eq:moment}) provides explicit expressions for arbitrary normally ordered moments, from which physically relevant observables such as field quadratures and higher-order correlation functions follow immediately. Because these expressions are obtained analytically, they eliminate the need for repeated numerical integration of the master equation while avoiding errors associated with finite Fock-space truncation. The analytical expressions agree with numerical simulations across a broad range of system parameters while providing a substantially more efficient method of evaluating expectation values.

Appendix~\ref{app:heisenberg_structure} concludes with a structural observation motivated by the present derivation. Incorporating coherent driving or continuous measurement introduces terms linear in the ladder operators, enlarging the underlying $\mathfrak{su}(1,1)\oplus\mathfrak{u}(1)$ algebra to a semidirect product with a Heisenberg algebra. While this observation does not by itself establish exact solvability for these more general systems, it suggests that the Lie-algebraic approach developed here may extend beyond the undriven master equation. Such systems arise naturally in continuously monitored quantum dynamics and driven Kerr platforms used for continuous-variable quantum information processing, making them promising directions for future investigation.
%============================================================
\appendix
\section{Derivation of the Wei--Norman Equations}
\label{app:WeiNorman}
%============================================================

The propagator is written in the disentangled form
\[
U(t)
=
e^{\Gamma_-(t)K_-}
e^{\Gamma_3(t)K_3},
\]
which satisfies
\[
\frac{dU}{dt}
=
\left(
\kappa K_-+\lambda K_3
\right)U,
\qquad
\lambda=-2i\chi K_0-\kappa.
\]
Differentiating the ansatz gives
\[
\dot U
=
\left(
\dot\Gamma_-K_-
+
e^{\Gamma_-K_-}
\dot\Gamma_3K_3
e^{-\Gamma_-K_-}
\right)U.
\]
Using the Baker--Campbell--Hausdorff identity together with
\[
[K_-,K_3]=-K_-,
\]
the nested commutator series terminates after the first term,
\[
e^{\Gamma_-K_-}
K_3
e^{-\Gamma_-K_-}
=
K_3-\Gamma_-K_-.
\]
Hence
\[
\dot U
=
\left[
(\dot\Gamma_--\Gamma_-\dot\Gamma_3)K_-
+
\dot\Gamma_3K_3
\right]U.
\]

Equating the coefficients of the linearly independent generators
$K_-$ and $K_3$ with those of the evolution equation yields
\[
\dot\Gamma_3=\lambda,
\qquad
\dot\Gamma_-+\lambda\Gamma_-=\kappa.
\]
\section{Proof of Combinatorial Identity}
\label{app:identity}

We prove the identity
\begin{equation}
\sum_{r=0}^{m}
a^r
\binom{m}{r}
\frac{(m-r)!}{(m-r-\ell)!}
=
\frac{m!}{(m-\ell)!}(1+a)^{m-\ell}
\end{equation}
for $0\le \ell \le m.$
\begin{proof}
Using
\[
\binom{m}{r}=\frac{m!}{r!(m-r)!},
\]
the left-hand side becomes
\[
m!\sum_{r=0}^{m}
\frac{a^r}{r!(m-r-\ell)!}.
\]
Since $(m-r-\ell)!$ is defined only for $r\le m-\ell$, the sum reduces to
\[
m!\sum_{r=0}^{m-\ell}
\frac{a^r}{r!(m-\ell-r)!}.
\]
Factoring out $(m-\ell)!$ gives
\[
\frac{m!}{(m-\ell)!}
\sum_{r=0}^{m-\ell}
\binom{m-\ell}{r}a^r,
\]
and the binomial theorem immediately yields
\[
\frac{m!}{(m-\ell)!}(1+a)^{m-\ell},
\]
which is the desired identity.
\end{proof}
\section{Semidirect Extensions by the Heisenberg Algebra}
\label{app:heisenberg_structure}

The damped Kerr Liouvillian considered in this work is generated entirely by the
$\mathfrak{su}(1,1)\oplus\mathfrak{u}(1)$ algebra. However, physically relevant extensions, such as coherently driven Kerr oscillators or stochastic master equations \cite{WisemanMilburn}, introduce terms linear in the left and right ladder operators
$a_{L}^{(\dagger)}$ and $a_{R}^{(\dagger)}$.

These operators generate the two-mode Heisenberg algebra
\[
\mathfrak{h}
=
\operatorname{span}\{a_L,a_L^\dagger,a_R,a_R^\dagger,\mathbb{I}\},
\]
with commutation relations
\[
[a_L,a_L^\dagger]=\mathbb{I},
\qquad
[a_R,a_R^\dagger]=\mathbb{I},
\qquad
[a_L^{(\dagger)},a_R^{(\dagger)}]=0.
\]
Hence
\[
[\mathfrak h,\mathfrak h]=\operatorname{span}\{\mathbb{I}\},
\qquad
[[\mathfrak h,\mathfrak h],\mathfrak h]=0,
\]
so $\mathfrak h$ is a two-step nilpotent Lie algebra.

The key observation is that the generators
\[
K_\pm,\quad K_3,\quad K_0
\]
act on $\mathfrak h$ by adjoint derivations:
\[
[K_i,X]\in \mathfrak h,
\qquad X\in \mathfrak h.
\]
Thus $\mathfrak h$ is stable under the adjoint action of
$\mathfrak{su}(1,1)\oplus\mathfrak{u}(1)$, forming a semidirect product
structure
\[
(\mathfrak{su}(1,1)\oplus\mathfrak{u}(1)) \ltimes \mathfrak h.
\]
This structure shows that adding linear driving or measurement terms enlarges
the algebra without generating higher-order polynomials in the ladder
operators. Consequently, the enlarged algebra may still be amenable to
Wei--Norman disentangling \cite{WeiNorman1963}, suggesting a possible extension of the present
approach to driven and continuously measured Kerr systems.
\bibliographystyle{apsrev4-2}
\bibliography{references}

@book{ScullyZubairy,
  title     = {Quantum Optics},
  author    = {Scully, M. O. and Zubairy, M. S.},
  year      = {1997},
  publisher = {Cambridge University Press}
}

@book{BreuerPetruccione,
  title     = {The Theory of Open Quantum Systems},
  author    = {Breuer, Heinz-Peter and Petruccione, Francesco},
  year      = {2002},
  publisher = {Oxford University Press}
}

@book{GerryKnight2005,
  title     = {Introductory Quantum Optics},
  author    = {Gerry, C. C. and Knight, P. L.},
  year      = {2005},
  publisher = {Cambridge University Press}
}

@book{WisemanMilburn,
  title     = {Quantum Measurement and Control},
  author    = {Wiseman, H. M. and Milburn, G. J.},
  year      = {2009},
  publisher = {Cambridge University Press}
}

@article{Lindblad1976,
  author  = {G{\"o}ran Lindblad},
  title   = {On the Generators of Quantum Dynamical Semigroups},
  journal = {Communications in Mathematical Physics},
  volume  = {48},
  pages   = {119--130},
  year    = {1976}
}

@article{Milburn1986SchrodingerCats,
  author  = {Milburn, G. J.},
  title   = {Quantum and classical Liouville dynamics of the Kerr medium},
  journal = {Physical Review A},
  volume  = {33},
  pages   = {674},
  year    = {1986}
}

@article{YurkeStoler1986,
  author  = {Yurke, B. and Stoler, D.},
  title   = {Generating quantum mechanical superpositions of macroscopically distinguishable states via the Kerr medium},
  journal = {Physical Review Letters},
  volume  = {57},
  pages   = {13--16},
  year    = {1986}
}

@article{Chaturvedi1991,
  author  = {Chaturvedi, S. and Srinivasan, V.},
  title   = {Solution of the master equation for an attenuated or amplified nonlinear oscillator with an arbitrary initial condition},
  journal = {Journal of Modern Optics},
  volume  = {38},
  pages   = {777--783},
  year    = {1991}
}

@article{BlaisReview,
  author  = {Blais, Alexandre and Girvin, S. M. and Oliver, W. D.},
  title   = {Circuit quantum electrodynamics},
  journal = {Reviews of Modern Physics},
  volume  = {93},
  pages   = {025005},
  year    = {2021}
}

@article{Krantz2019,
  author  = {Philip Krantz and Morten Kjaergaard and Fei Yan and Terry P. Orlando and Simon Gustavsson and William D. Oliver},
  title   = {A Quantum Engineer's Guide to Superconducting Qubits},
  journal = {Applied Physics Reviews},
  volume  = {6},
  number  = {2},
  pages   = {021318},
  year    = {2019},
  doi     = {10.1063/1.5089550}
}

@article{WeiNorman1963,
  author  = {Wei, James and Norman, Eugene},
  title   = {Lie Algebraic Solution of Linear Differential Equations},
  journal = {Journal of Mathematical Physics},
  volume  = {4},
  number  = {4},
  pages   = {575--581},
  year    = {1963},
  doi     = {10.1063/1.1703993}
}

\end{document}